\documentclass[conference]{IEEEtran}
\usepackage{cite}
\usepackage{amsmath,amssymb,amsfonts}
\usepackage{algorithmic}
\usepackage{graphicx}
\usepackage{textcomp}
\usepackage{xcolor}
\usepackage{tabularx}
\usepackage{array}
\usepackage[hyphens]{url}
\usepackage[font={small}]{caption}
\usepackage{subcaption}
\def\BibTeX{{\rm B\kern-.05em{\sc i\kern-.025em b}\kern-.08em
    T\kern-.1667em\lower.7ex\hbox{E}\kern-.125emX}}
\begin{document}

\title{Generative AI Beyond LLMs: System Implications of Multi-Modal Generation}

\author{Alicia Golden$^{1,2}$ \enspace Samuel Hsia$^{1,2}$ \enspace Fei Sun$^{3}$ \enspace Bilge Acun$^{1}$ \enspace Basil Hosmer$^{1}$ \enspace Yejin Lee$^{1}$\\ 
Zachary DeVito$^{1}$ \enspace Jeff Johnson$^{1}$ \enspace Gu-Yeon Wei$^{2}$ \enspace David Brooks$^{2}$ \enspace Carole-Jean Wu$^{1}$ 
\\ \\
$^{1}$FAIR at Meta \quad $^{2}$Harvard University \quad $^{3}$Meta
}


\maketitle

\begin{abstract}
As the development of large-scale Generative AI models evolve beyond text (1D) generation to include image (2D) and video (3D) generation, processing spatial and temporal information presents unique challenges to quality, performance, and efficiency. We present the first work towards understanding this new system design space for multi-modal text-to-image (TTI) and text-to-video (TTV) generation models. Current model architecture designs are bifurcated into 2 categories: Diffusion- and Transformer-based models. Our systematic performance characterization on a suite of eight representative TTI/TTV models shows that after state-of-the-art optimization techniques such as Flash Attention are applied, \textit{Convolution} accounts for up to 44\% of execution time for Diffusion-based TTI models, while \textit{Linear} layers consume up to 49\% of execution time for Transformer-based models. We additionally observe that Diffusion-based TTI models resemble the {\tt Prefill} stage of LLM inference, and benefit from 1.1-2.5x greater speedup from Flash Attention than Transformer-based TTI models that resemble the {\tt Decode} phase. Since optimizations designed for LLMs do not map directly onto TTI/TTV models, we must conduct a thorough characterization of these workloads to gain insights for new optimization opportunities. In doing so, we define sequence length in the context of TTI/TTV models and observe sequence length can vary up to 4x in Diffusion model inference. We additionally observe temporal aspects of TTV workloads pose unique system bottlenecks, with Temporal Attention accounting for over 60\% of total Attention time. Overall, our in-depth system performance characterization is a critical first step towards designing efficient and deployable systems for emerging TTI/TTV workloads.
\end{abstract}

\begin{IEEEkeywords}
Generative AI, Multi-Modal, Diffusion Model, Transformer, Sequence Length, Attention
\end{IEEEkeywords}

\section{Introduction}
\label{section1}

Recent advancements in generative AI have prompted immense effort into the development of efficient and scalable models for text generation \cite{efficienttransformerinference, efficienttransformer2, efftrans}. The advent of Large Language Models (LLMs) has spurred myriad applications including chatbots, such as ChatGPT \cite{openai2023chatgpt}, email assistants \cite{ai2}, and coding tools \cite{github_copiolet}. Significant effort has gone into increasing the efficiency of these models when deployed at-scale, helping enable ChatGPT alone to serve over 100 million active users per week \cite{100mill}. Yet text generation is just the tip of the iceberg. A one-dimensional representation of information lacks the ability to convey spatial and temporal information, both of which are critical for understanding the world around us. The natural progression of these large-scale generative AI models is thus to evolve from \textit{text} (1D) to \textit{image} (2D) to \textit{video} (3D). However, moving to higher dimensional representations presents numerous challenges to quality, performance, and efficiency. While current systems have mainly been optimized for LLMs via techniques such as Flash Attention \cite{flashattention}, the distinct properties of Text-To-Image (TTI) and Text-To-Video (TTV) models suggest these emerging workloads may not see equal benefits --- thus requiring an in-depth analysis to identify opportunities for TTI/TTV optimization. This paper carefully examines emerging TTI/TTV workloads, and highlights attributes that greatly differ from predecessor text-based LLMs.

While image and video generation models have seen significant \textit{algorithm} advancements in recent years, relatively little has been done to optimize the deployment of these models from a \textit{systems} perspective. New system optimizations tailored towards system performance bottleneck of TTI/TTV models have the potential to replace the generation of short video clips on the order of seconds with full movies and films. In this paper, we refer to the \textit{performance bottleneck} as operators that dominate the execution time of the workload. Other image generation tasks such as sticker-generation \cite{sticker_generation}, educational material \cite{education}, and even scientific discoveries \cite{mattergen}, could likewise benefit from system optimizations that enable increased speed and resolution. Overcoming systems challenges is critical to enabling future applications.

 \begin{figure} 
     \centering
     
         \includegraphics[width=\linewidth]{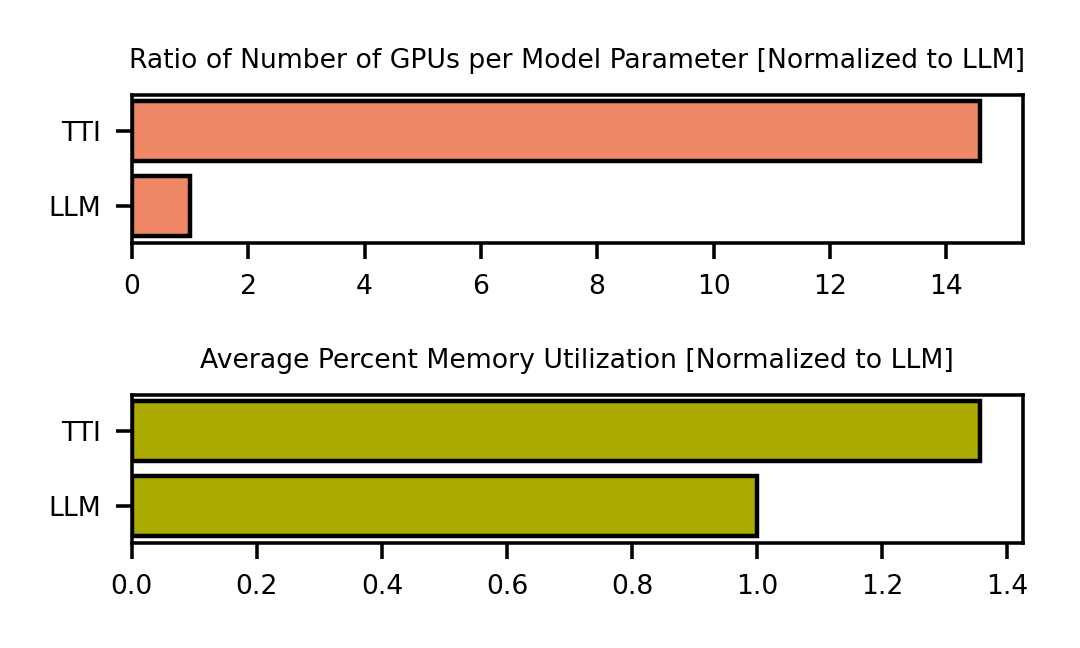}
     
     \caption{Across industry-scale datacenters, Text-to-Image (TTI) models use roughly 14x more GPUs per model parameter during training and 1.35x higher memory utilization as compared to LLMs, demonstrating their growing importance at the datacenter scale. }
    \label{fig:fleetwide}
\end{figure}

To evaluate current state-of-the-art image/text generation, we first examine industry-scale generative deep learning tasks at the fleet-wide level. We find that while TTI/TTV models are \textit{an order of magnitude smaller} than LLMs in terms of model parameters, the number of GPUs used for training is roughly in the same order-of-magnitude. In fact, the ratio of number of GPUs per model parameter is 14x higher for TTI models than LLMs, emphasizing the importance of running these models efficiently (Figure \ref{fig:fleetwide}). Further fleet-wide characterization  reveals that this emerging class of AI workloads has distinct system requirements --- average memory utilization for TTI/TTV models is roughly 35\% higher than LLMs.

We subsequently take a quantitative approach to characterizing state-of-the-art TTI/TTV models, comparing the multi-dimensional design space across latency and computational intensity. We construct a model suite of eight representative text-to-image and video generation tasks and demonstrate how these models are distinct from widely-used language models, i.e., LLaMA \cite{llama2}. We find that new system performance bottlenecks emerge after the application of state-of-the-art performance optimizations, such as Flash Attention \cite{flashattention}, with \textit{Convolution} accounting for up to 44\% of execution time in Diffusion-based TTI models, and \textit{Linear} layers consuming up to 49\% of execution time in Transformer-based TTI models. 

We additionally observe that traditional LLM paradigms such as \textit{Prefill/Decode} and \textit{Sequence Length}, do not map 1:1 onto TTI/TTV workloads. We profile sequence length over the course of inference and find that in contrast to LLMs, sequence length varies by up to 4x over the course of Diffusion model inference. In general, sequence length scales quadratically with image size in Diffusion models. Furthermore, we investigate the system performance bottleneck presented by Temporal Attention, which allows for cohesive frame generation across time in TTV models. This is in contrast to Spatial Attention, which attends across a 2D image.  We find that Temporal Attention takes 2x the execution time as Spatial Attention, yet consumes 9x the FLOP count. We further observe the Temporal Attention bottleneck grows quadratically with number of frames, suggesting the need for future optimizations.

\begin{figure*}
    \centering
    \includegraphics[width=0.9\linewidth]{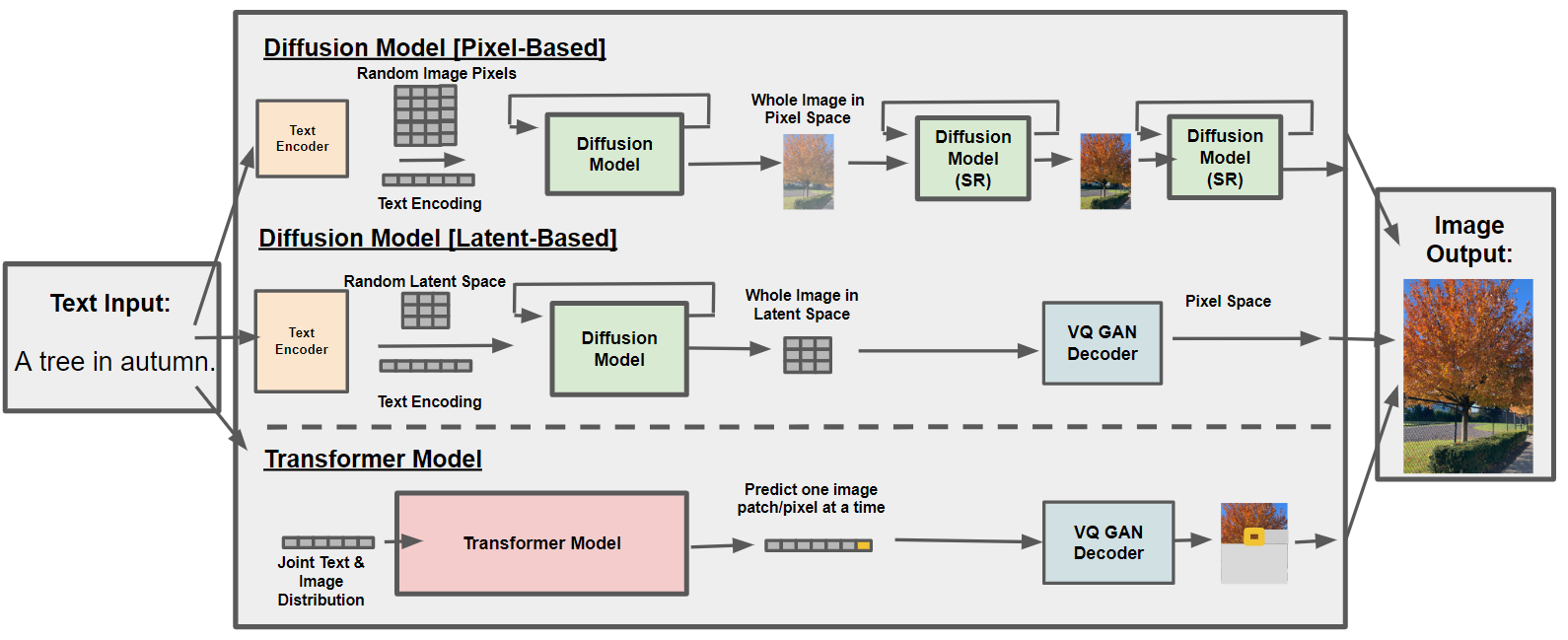}
    \caption{Common Text-to-Image Model Architectures. Models consist of multiple independently-trained components, and are strung together during inference (shown here) to take text as input and generate an image output. Note that the top two models use a diffusion-based architectures (green), while bottom two models use transformer-based architectures (red).}
    \label{fig:tti_architecture}
\end{figure*}

Our subsequent analysis on industry-scale generative AI use cases provides interesting insights for future system design. \textbf{TTI/TTV models exhibit the following unique system properties, which differ from traditional LLMs:}

\begin{itemize}
    \item \textbf{High arithmetic intensity (Section \ref{section2}).}  TTI models, and in particular diffusion-based models exhibit higher arithmetic intensity as compared to LLMs by up to 100x. This stems from the high parameter reuse that occurs during the UNet, where tens or hundreds of denoising steps cause iterations over the same number of parameters.
    
    \item \textbf{Convolution as a system performance bottleneck (Section \ref{section4}-A).} After state-of-the-art optimizations such as Flash Attention are applied, the system performance bottleneck shifts to other operators such as \textit{Convolution}, which accounts for up to 44\% of execution time.

    \item \textbf{Unique prefill/decode correspondence (Section \ref{section4}-B)}. Diffusion model inference resembles the \textit{Prefill} phase of LLM inference, whereas Transformer-based TTI models resemble the \textit{Decode} phase. We find that Attention kernel speedup when using Flash Attention is 1.1-2.5x greater for Diffusion models as compared to Transformer-based TTI models. This suggests that Diffusion- and Transformer-based TTI models require different optimization techniques.

    \item  \textbf{Highly variable sequence length (Section \ref{section5}-A). } We profile sequence length over the course of inference and find sequence lengths in Diffusion models such as Stable Diffusion can vary by up to 4x. This variable sequence length impacts the computational intensity over the course of Diffusion model inference, and poses opportunities to tailor system design.

    \item  \textbf{Scaling dependence with image size (Section \ref{section5}-B).} We find that Attention memory requirements scale as \math O(L^4) \endmath, where L is image/latent dimension. Larger image sizes see significantly higher memory requirements. Furthermore, as we increase image size to higher resolutions, we find that Convolution execution time in Diffusion models scales faster than Attention after state-of-the-art techniques such as Flash Attention are applied.

    \item \textbf{Temporal dimension (Section \ref{section6})}. The temporal dimension inherent in TTV models presents a new system bottleneck as compared to the TTI models from which they are built. Through profiling, we find that Temporal Attention suffers from 2x slower execution time as compared to Spatial Attention, even as it requires 9x the FLOP count. This suggests temporal attention is an important bottleneck for system design.

\end{itemize}

\section{Understanding Multi-Modal Machine Learning Tasks} \label{section2}

We first present a system-informed taxonomy of the current landscape of text-to-image (TTI) and text-to-video (TTV) generation models. Recent model development has seen a bifurcation of architectures, each characterized by their own image generation process and system characteristics. Figure~\ref{fig:tti_architecture} illustrates the image generation pipeline for these two classes of workloads: (i) Diffusion Models, including both \textit{pixel-} and \textit{latent-} based models, and (ii) Transformer-based Models. These same two classes of workloads additionally form the fundamental building blocks of TTV models --- since video generation models typically generate a series of images (i.e., frames) using a pretrained TTI model, and then ensure the frames are temporally consistent through additional temporal attention/convolutional layers. The subsequent discussion and analysis refers to the \textit{Inference} pipeline of image/text generation. Unlike LLMs, TTI/TTV models consist of several different model components that are trained separately and then stiched together at inference time.

\subsection{Text-to-Image Generation Models}

\subsubsection{Diffusion Models}

Diffusion models generate images in an iterative \textit{denoising} process, where a random group of pixels is gradually transformed into an image that resembles a given text input \cite{denoising1, denoising2, denoising3}. During the image generation pipeline (Figure \ref{fig:tti_architecture}), a text input is first encoded into an embedding representation before being passed into the diffusion model, along with a set of random noise pixels. As shown in Figure 3, the diffusion model architecture follows a \textit{UNet} structure, which gradually downsamples and upsamples a noisy image through alternating \textit{Attention} and \textit{Resnet} blocks of various sizes \cite{unet, stablediffusion}. Note the Resnet blocks alternate with (i) \textbf{Self-Attention} blocks, which condition the generation on the image itself, and (ii) \textbf{Cross-Attention}, which attends to the text input. Before generating a final output, the image traverses through the UNet tens or hundreds of times as part of the denoising process, leading to high compute intensity, frequent parameter reuse, and long latencies (see Section \ref{section2c}). Note there is an inherent trade off between number of denoising steps and image quality. 

As shown in Figure \ref{fig:tti_architecture}, there are two distinct variations of diffusion models from a systems perspective --- \textit{pixel} and \textit{latent} based. Pixel vs Latent models are distinguished by the parameter space of the diffusion model and the subsequent post-processing which is necessary once an initial image is generated. While pixel-based models operate the denoising process on standard image pixels, latent-based models transform the image into an embedding representation, making it more efficient for computation \cite{stablediffusion}. As a result, latent-based models can represent high-resolution images without the need to feed the image through subsequent SuperResolution (SR) networks as in the pixel case. This comes at the cost of a VAE or GAN-based decoder to convert latent space back to pixel space once finished.

\subsubsection{Transformer Models}

\begin{figure}
     \centering
     
     \begin{subfigure}[t]{\linewidth}
         \centering
         
         \includegraphics[width=\linewidth]{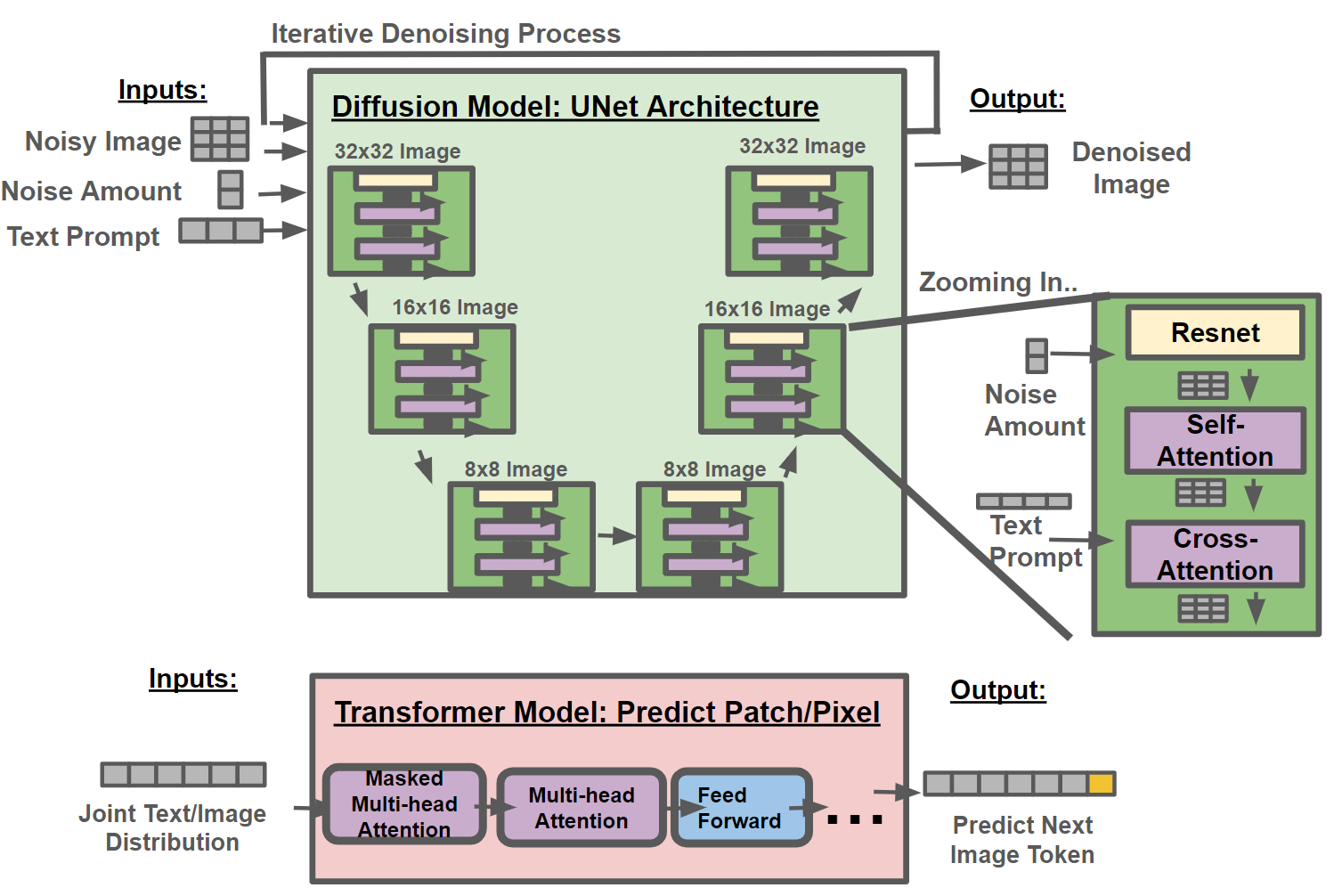}
         
         \label{fig:temp2}
     \end{subfigure}%
     \caption{Detail on Diffusion and Transformer models. Note that Diffusion models consist of Resnet blocks, Self-Attention blocks, and Cross-Attention blocks while Transformer-based models consider Self/Cross Attention and FeedForward. }
    \label{fig:detailed_architectures}
\end{figure}

In contrast to the iterative nature of Diffusion models, Transformer-based TTI models generate an image sequentially. As shown in Figure~\ref{fig:tti_architecture}, the transformer-based architectures model image generation as a sequence-to-sequence task, where the prediction of the next pixel (or image patch) is conditioned on all the previous patches \cite{dalle}. Note that image tokens generated from the transformer model must then be decoded through a GAN decoder or equivalent to convert back to an image representation. Figure \ref{fig:detailed_architectures} shows a detailed view of the basic transformer architecture, which consists of two multi-headed attention layers and a feedforward layer, and remains unchanged from LLMs. However, the number and arrangement of these transformer blocks varies. Compared to GPT-3, which has 96 layers and a model dimension of 12,288, Parti has 80 layers and a model dimension of 4096  \cite{parti, gpt3}. Other transformer-based TTI models such as Muse have 48 layers and a hidden dimension of 2048 \cite{muse}.

\subsection{Text-to-Video Models}

Text-to-Video models extend traditional TTI model architectures, and often use pretrained TTI models to generate individual image frames before connecting these together with temporal attention or convolution. Like TTI models, TTV models can follow a diffusion-based~\cite{singer2022makeavideo} or a transformer-based model structure \cite{phenaki}. However, creating a temporally-cohesive video is challenging from a systems perspective. For example, \textbf{Temporal Attention} layers are often inserted after existing \textbf{Spatial Attention} layers in the UNet architecture (Figure \ref{fig:detailed_architectures}), since adding an additional temporal dimension to the existing Attention call is not feasible from a memory perspective. Additionally, Attention calls are sometimes substituted for Convolutional layers to keep computational/memory costs down, especially in models with higher resolution \cite{imagenvideo}.

\begin{figure}[t!]
    \centering
    
    \includegraphics[width=0.93\linewidth]{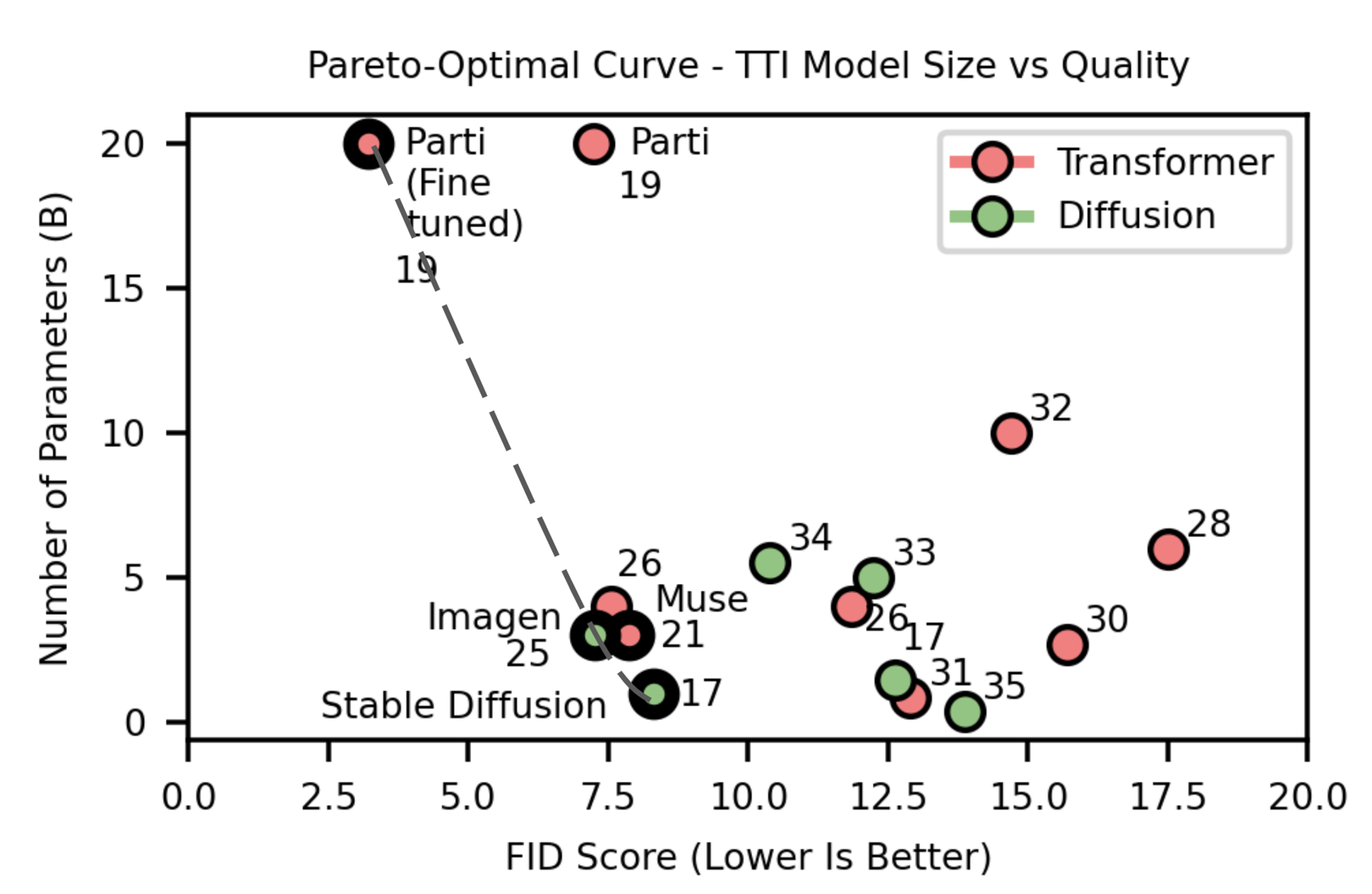}
    \caption{Pareto-Optimal curve showing tradeoff between model quality and system resources  for various Text-to-Image models. Bottom left corner is optimal. Bolded points represent models further examined in model suite outlined in Section \ref{section3}. Note that a corresponding figure for TTV models shows a similar trend. Diffusion TTV models often have both the lowest number of parameters and FID score.}
    \label{fig:taxonomy}
\end{figure}

\newcolumntype{b}{X}
\newcolumntype{s}{>{\hsize=2\hsize}X}
\newcolumntype{e}{>{\hsize=1.2\hsize}X}
\newcolumntype{f}{>{\hsize=0.8\hsize}X}

\begin{table} [b]
\begin{center}{
\small
\renewcommand\tabularxcolumn[1]{>{\centering\arraybackslash}m{#1}}
\begin{tabularx}{\linewidth}{|s|b|e|f|b|b|}
 \hline
 Model & Imagen \cite{imagen} & Stable Diffusion\cite{stablediffusion} & Muse \cite{muse} & Parti \cite{parti}\\ 
 \hline\hline
 Architecture & Diffusion (Pixel) & Diffusion (Latent) &Trans- former &Trans- former \\ 
 \hline
Num Params & 3B & 1.45B & 3B & 20B  \\
 \hline
Num Layers & --- & --- & 48 & 80 \\
\hline
Model Dim & --- & --- & 2048 & 4096  \\
\hline
Attn Res & [32,16,8] & [4, 2, 1] & --- & ---  \\
\hline
Text Cross & [32,16,8] & --- & --- & ---  \\
Attn Res & &  &  &  \\
\hline
Channel Mult &[1,2,4,4] & [1,2,4,4] &  &  \\
\hline
Num Res Blocks & 3 & 2 & --- & ---  \\
\hline
Per-Head Channels & 64 & 8 & --- & ---  \\
\hline
Embed Dim & 512 & 768 & --- & ---  \\
\hline\hline
Compute & High & Medium & Low & Low \\
\hline
Memory & Medium & Low & Low & High  \\
\hline
Latency & High & Medium & Low  & Medium  \\
\hline
\end{tabularx}}
\end{center}
\caption{Taxonomy of Text-to-Image Models}
\label{tbl:taxonomy_table}
\end{table}

\begin{figure}
    
     \includegraphics[width=\linewidth]{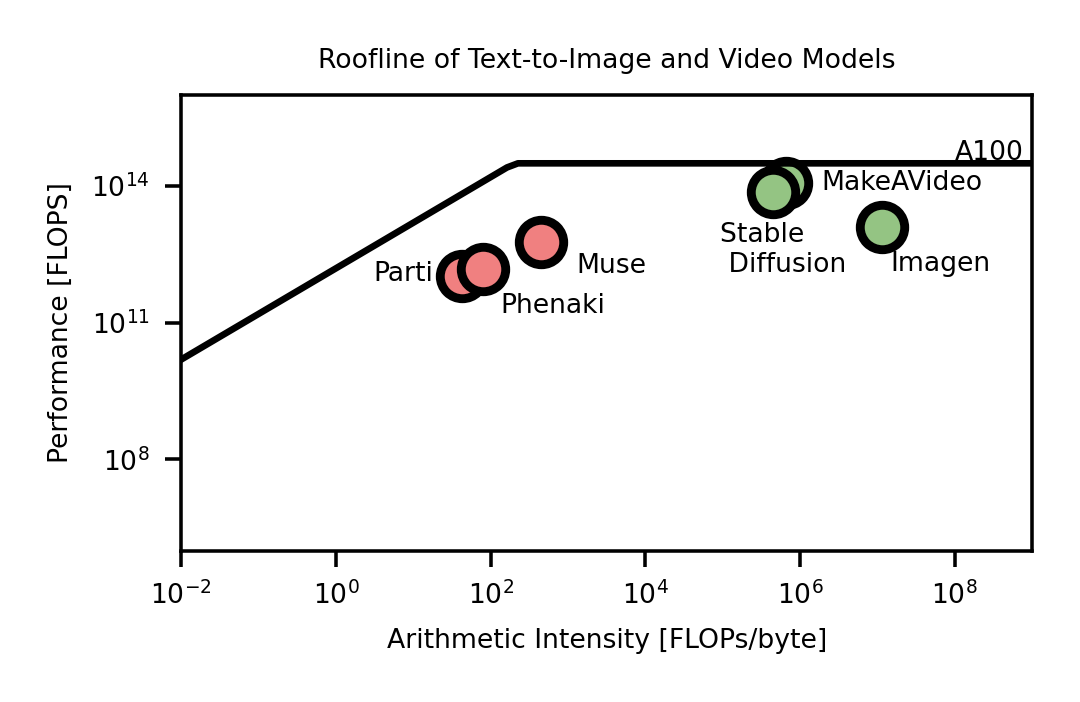}
    \caption{Text-to-Image/Video Models Roofline on A100 GPU. Diffusion models have higher arithmetic intensity than transformer-based TTI models, and fall in the compute-bound region. Transformer-based models are memory-bandwidth bound at low batch sizes.}
    \label{fig:tti_roofline}
\end{figure}

\subsection{System Design Space of Text-to-Image/Video Models} \label{section2c}

To further understand the system design space for emerging multi-modal generative AI technologies, Figure~\ref{fig:taxonomy} illustrates state-of-the-art TTI generation technologies between the key design dimensions of model accuracy (x-axis) and the number of trainable parameters (y-axis). Models are labeled as their citation number \cite{dalle, makeascene, cogview, cogview2, cm3, racm3, muse, parti, nuwa, ernievilg, glide, dalle2, stablediffusion, imagen, vqdiffusionf}. Accuracy is measured via previously reported values of FID score \cite{fid_score} on the COCO dataset \cite{cocodataset}. We omit non-open source models since the lack of publicly available implementations prevents a more in-depth system analysis. As shown, the most optimal models tend towards the bottom left corner with lowest FID score and fewest number of parameters. In general, Diffusion models tend to have higher model quality for the same number of parameters as Transformer models. However, a diverse set of models lie on the Pareto-Optimal curve, including Imagen~\cite{imagen} (pixel-based), Stable Diffusion~\cite{stablediffusion} (latent based) and Parti (transformer), the last of which offers higher model quality than diffusion models but at 4x the parameter count. This exemplifies the importance of understanding system implications for these various architectures.

Moving beyond parameter count, we next categorize these model architectures along the major system design axes of \textit{compute}, \textit{memory}, and \textit{latency}. Table \ref{tbl:taxonomy_table} highlights specs and system requirements of four of the models from the Pareto-Optimal curve shown in Figure \ref{fig:taxonomy}, while Figure \ref{fig:tti_roofline} plots these models on a roofline corresponding to an A100 GPU. Note arithmetic intensity is calculated as ratio of FLOPs to required model capacity.  We first find that Diffusion models have a higher arithmetic intensity than Transformer-based TTI models. This is due to the \textit{denoising} process inherent to Diffusion models. The large number of iterations through the UNet incurs high FLOP count on a small number of parameters, leading to high parameter re-use. The denosing process also incurs high latency, given the large number of iterations required. In comparison, Transformer-based TTI models tend to be memory bound for the low batch size case shown in Figure \ref{fig:tti_roofline}. Low batch sizes are appropriate for TTI models. Transformer-based TTI require less compute and often higher memory requirements, especially in the case of Parti~\cite{parti}. Yet transformer TTI models in general have faster latencies as compared to the iterative diffusion process. This is especially true with recently introduced techniques such as parallel decoding \cite{muse}.

\section{Experimental Methodology} \label{section3}

We construct a model suite of eight workloads representative of the model architectures introduced in Section \ref{section2}, including comparing against a state-of-the-art, publicly-available text-to-text generation model --- LLaMA2 \cite{llama2}. In addition to the four open-source models highlighted in Section \ref{section2}, we further augment our model suite to provide a realistic view of system requirements for deployment at-scale by including a production TTI model.  We evaluate all models on real system hardware and measure their system requirements.

\textit{Workloads. }
To construct the model suite, we select representative models for each category of text-to-image model architecture (Section \ref{section2}). Specifically, we include models that are on the Pareto-Optimal Curve between model quality and number of parameters (see Figure \ref{fig:taxonomy}). We select Imagen as a representative pixel-based diffusion model, given its presence on the Pareto Optimal curve. Imagen contains a T5 Encoder to encode textual information and includes three diffusion models: one to produce a 64x64 image, and two other super-resolution models that serve to upsample the image to 768x768 and 1024x1024, respectively. Additionally, to represent a latent-based model we select a model using Stable Diffusion architecture retrained on licensed data, which includes a CLIP text encoder, diffusion model, and a VQ VGAN model. For transformer-based TTI models, we select Muse and Parti to showcase the diversity of these transformer architectures, as Muse is a decoder-only transformer model that uses parallel decoding at inference time, while Parti is an encoder-decoder model that predicts image tokens autoregressively. We also include two TTV models: \texttt{Make-a-Video}~\cite{singer2022makeavideo} is built upon a diffusion model architecture, and \texttt{Phenaki}~\cite{phenaki} that is derived from transformers.

\textit{Hardware Systems. } We evaluate training and inference using NVIDIA A100 80GB GPUs. For inference analysis, we profile TTI/TTV model inference on a single GPU, since the model parameters can fit within the 80 GB memory constraints. When profiling model training, we use Fully Sharded Data Parallelism (FSDP) to train over multiple compute nodes, where each node consists of 8 A100 GPUs. Figure \ref{fig:compute_comm_scaling} shows how compute and communication scale as training is distributed over an increasing number of nodes.

\textit{Tools. } To perform our subsequent characterization and analysis, we use PyTorch Profiler to record execution timelines. We measure GPU kernel time and annotate the model to track which CPU operators correspond to a given GPU launch kernel. We develop a profiling framework to automate this process, via inserting hooks into the forward pass. We then develop scripts to efficiently parse the resulting Pytorch Profiler output and link GPU kernels to their corresponding annotation to determine operator breakdowns, speedup, etc. We construct a similar framework to analytically calculate FLOPs. We use the NVIDIA Nsight Compute tool to examine kernel breakdowns and analyze cache hit rates, memory utilization, etc. 

\begin{figure} [t]
     \includegraphics[width=\linewidth]{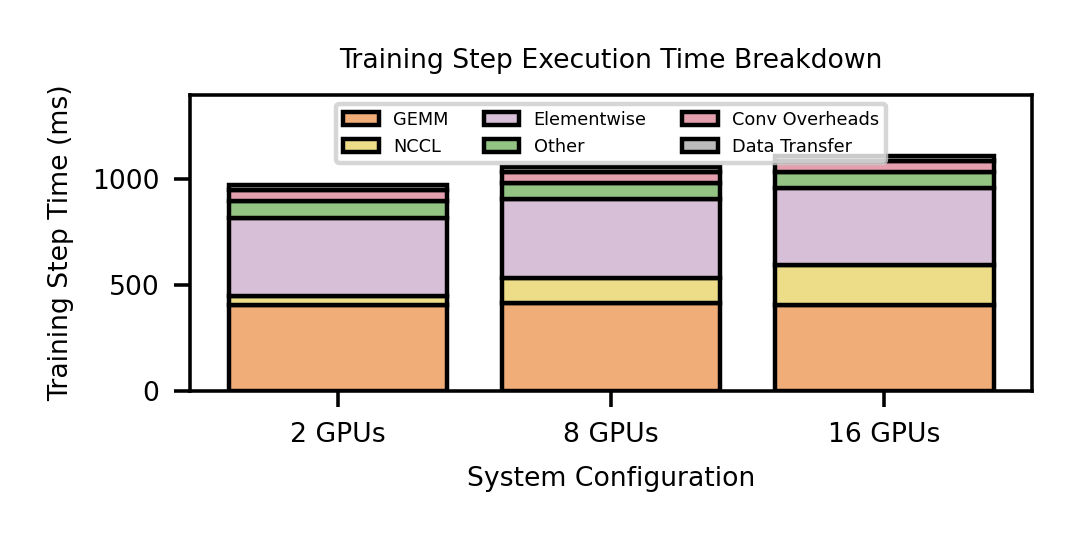}
    \caption{Compute versus communication scaling of Stable Diffusion model as training is distributed over an increasing number of nodes (i.e., 2, 8, and 16 A100 GPUs).}
    \label{fig:compute_comm_scaling}
\end{figure}
\section{System Performance Characterization} \label{section4}

\begin{figure*}
     \centering

\begin{center}{
\small
\renewcommand{\figurename}{TABLE}
\renewcommand{\thefigure}{\Roman{figure}}
\setcounter{figure}{1}
\caption{End-to-end speedup of Flash Attention as compared to Baseline Attention}
\label{figure}
\setcounter{figure}{6}   
\begin{tabular}{|c|c| c| c | c | c | c|c|}
 \hline
   \ \  \ LLaMA  \  & \ \ \ Imagen \ \ \ & StableDiffusion & \ \ \ Muse \ \ \ & \ \ \ Parti \ \ \ & Prod Image & MakeAVideo & Phenaki \\
  \hline 
  1.52x &  1.22x & 1.67x & 1.11x & 1.17x & 1.04x & 1.06x & 1.15x  \\
\hline
\end{tabular}}
\end{center}
\label{tbl:attn_table}
\begin{subfigure}[t!]{\textwidth}
         \centering
       
           \includegraphics[width=\textwidth]{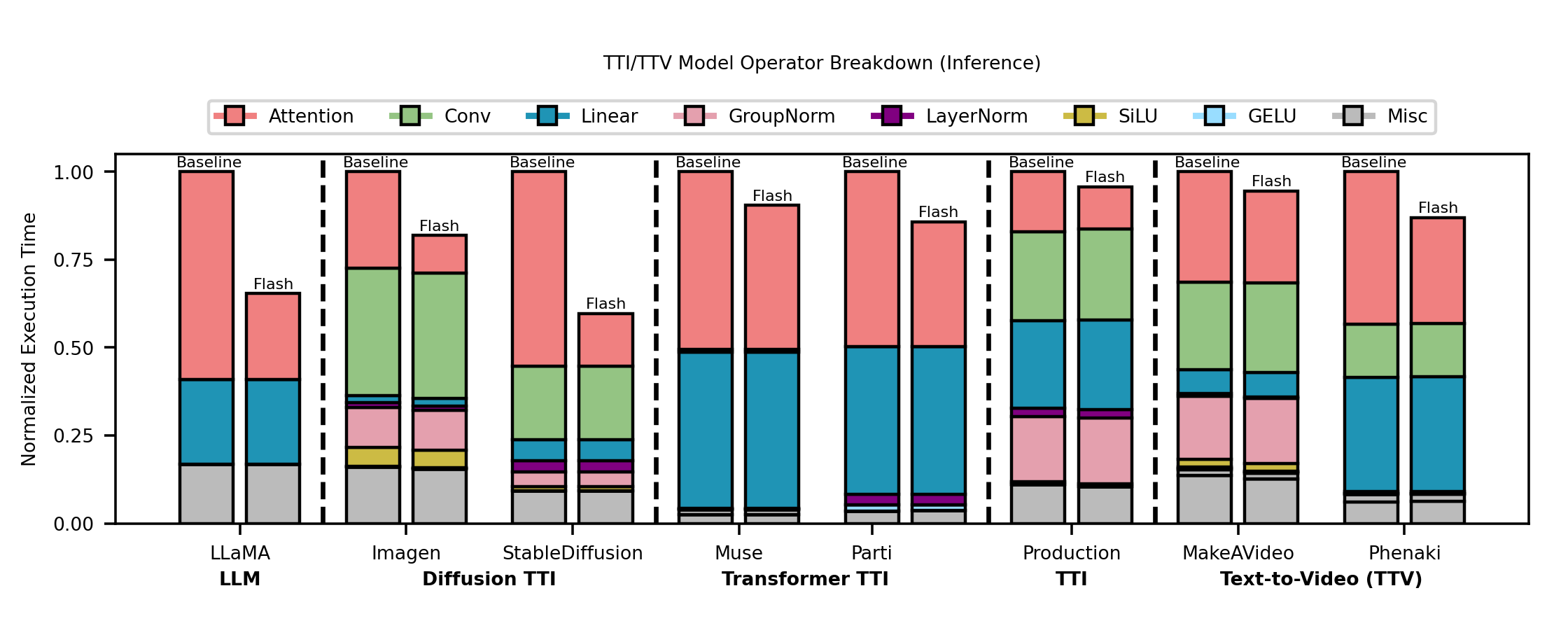}
     \end{subfigure}%
     \caption{Operator Breakdown Across TTI and TTV Models With Baseline Attention. First bar of each model shows model execution time with Baseline Attention, while second bar shows corresponding normalized execution time with Flash Attention.}
    \label{fig:model_suite}
\end{figure*}

Figure \ref{fig:model_suite} shows the operator time breakdown (y-axis) across the model suite that was introduced in Section~\ref{section3}.  On the y-axis, we compare the execution time of the forward pass (inference) pipeline shown in Figure \ref{fig:tti_architecture}. We record the execution time breakdown of the model using baseline Attention (left bar), and plot the  corresponding normalized execution time after Flash Attention V2 is applied (right bar).

\subsection{Analyzing System Performance Bottlenecks}
\label{section4a}
We first examine the execution time breakdown of baseline models (left). We observe the diversity of model operators in Diffusion-based TTI/TTV models as compared to traditional LLMs. While LLMs and transformer-based TTI models consist mainly of \textit{Attention} and \textit{Linear} layers, \textit{Convolution} accounts for up to 36\% of time in baseline Diffusion models. Diffusion models also have a larger variety of operators, with 4-11\% of execution time attributed to \textit{GroupNorm}. We additionally observe that \textit{pixel-based} models spend 15\% more time on convolution as compared to \textit{latent-based} models. This is due to a higher frequency of convolution operators, since pixel-based models contain super-resolution (SR) networks that follow a UNet architecture (Figure \ref{fig:tti_architecture}), but often swap attention layers for convolution due to prohibitive memory requirements at high resolutions. 

The detailed operator time characterization reveals Attention is an important system performance bottleneck in baseline models --- given that Attention consumes roughly 41.3\% of execution time when averaged across the TTI/TTV model suite. To accelerate the execution time of Attention, the recently proposed \textit{Flash Attention} shows significant performance potential on GPUs~\cite{flashattention}. The technique essentially allows for the large intermediate matrices produced in the Attention calculation to be tiled, thereby reducing memory accesses and offering significant speedup. Note that while Flash Attention was originally designed as a training optimization, previous work including the DeepSpeed Inference Framework and others has shown its benefit for inference as well~\cite{deepspeed}. Here we examine the impact of applying Flash Attention V2 across TTI/TTV inference workloads. 

Figure \ref{fig:model_suite} shows the resulting operator breakdown after Flash Attention V2 is applied, normalized to a given model's baseline execution time. Note that after Flash Attention is applied to LLaMA or transformer-based TTI models, Attention still accounts for 37-45\% of total execution time. In contrast, for Diffusion models, Attention consumes only 13-25\% of execution time after Flash Attention is applied, with Convolution remaining as the largest single operator block. \textbf{We find that after applying Flash Attention to Diffusion Models, the system performance bottleneck shifts to other operators such as \textit{Convolution} instead of Attention.}

\subsection{Evaluating Effectiveness of Flash Attention Across TTI/TTV Model Suite}
\label{prefilldecode}
\begin{table}
\small
\renewcommand\tabularxcolumn[1]{>{\centering\arraybackslash}m{#1}}
\begin{tabularx} {\columnwidth}{| >{\hsize=0.15\hsize}X | 
    >{\hsize=0.1\hsize}X | 
    >{\hsize=0.25\hsize}X | 
    >{\hsize=0.4\hsize}X |}
 \hline
   & \textbf{LLMs} & \textbf{Diffusion-Based Models} & \textbf{Transformer-Based Models} \\
 \hline
 \textbf{Training/ Prefill} & 1st token & Generate all pixels of the image at once & Process text prompt \\
 \hline
 \textbf{Decode} &  2nd token & N/A &  Generate each token autoregressively \\
  \hline 
\end{tabularx}
\setcounter{table}{2}
\caption{Prefill/Decode for LLMs versus TTI models}
\label{tbl:attn_table}
\end{table}

We additionally observe that the end-to-end speedup of Flash Attention varies from 4-67\% across the model suite. According to Amdahl's Law, the potential speedup such an optimization is impacted by two factors: (i) percent of time dedicated to Attention, and (ii) speedup of Attention module itself. While percent of time dedicated to Attention varies across model architectures, as illustrated in Section \ref{section4a}, here we focus on Attention module speedup. By examining Figure \ref{fig:model_suite} and comparing the isolated speedup of the Attention module (i.e., red bar), we find that Attention module speedup from Flash Attention is 1.1-2.5x greater for Diffusion Models as compared to Transformer-based TTI models. 

To understand the observed difference in Attention speedups across the model suite, we note that Attention kernel speedup varies as a factor of matrix size. We subsequently analyze TTI model speedup in the context of traditional LLM inference, which consists of two distinct phases: \textit{Prefill} and \textit{Decode}. \textit{Prefill} (i.e., the initial processing of the prompt) allows for greater parallelization, as it is characterized by the processing of a large Nxd query vector, where N is sequence length and d is model dimension. A large NxN similarity matrix is created during the Attention calculation, which benefits from the tiling of Flash Attention due to reduced HBM memory accesses \cite{flashattention}. In contrast, the \textit{Decode} stage generates tokens autoregressively, with queries of size 1xN. This generates a smaller Nx1 similarity matrix, which requires fewer memory accesses in the first place, and thus sees smaller speedup. Broadly speaking, \textit{Prefill} boils down to large matrix-matrix computations while \textit{Decode} consists of matrix-vector computations.

Table \ref{tbl:attn_table} illustrates how this concept of Prefill and Decode for LLM inference map onto TTI workloads. For diffusion-based TTI models, all pixels of an image are generated at once (see Figure \ref{fig:tti_architecture}), thus creating large matrices that resemble the prefill stage of LLM inference. This indicates Flash Attention is beneficial for Diffusion model inference. In contrast, image pixels (patches) are predicted sequentially in transformer-based TTI models due to their autoregressive nature, which resembles decoding. Note also that transformer TTI models see less speedup from Flash Attention as compared to LLMs, due to their smaller sequence length and matrix sizes (as discussed in Section \ref{section5}).
\textbf{Since traditional LLM paradigms such as prefill/decode do not apply in TTI/TTV workloads, this prompts a need to understand model features in the context of TTI/TTV generation in order to design optimal systems.}

\section{Impact of Sequence Length} \label{section5}

As illustrated by the varying effectiveness of Flash Attention, traditional LLM paradigms such as prefill/decode do not apply to Diffusion-based TTI and TTV workloads. In order to understand how these TTI/TTV models operate, we must translate other LLM concepts, such as sequence length, into the context of image/video generation. This will allow for more efficient system design.  

LLMs are characterized by a sequence that represents the information a model can attend to, i.e., the amount of words it can refer to when generating the next subsequent word \cite{sequence_length}. However, sequence length in state-of-the-art TTI/TTV models is directly impacted by the \textit{image size}. In particular, the sequence length of diffusion models is proportional to \math{(image\ size)^2} \endmath, as attention is computed between one version of the image and the next. In transformer-based TTI models, sequence length is impacted by \textit{image size} and text embedding size jointly. 

\subsection{Sequence Length Profiling}

\begin{figure*}
    \centering
   
    \includegraphics[width=0.8\linewidth]{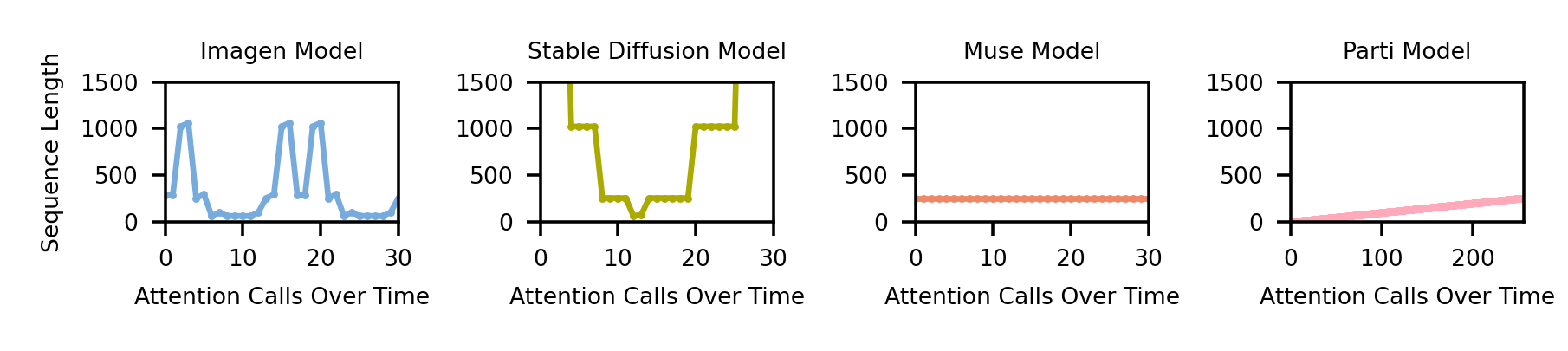}
    \caption{Sequence length profiling across various models in model suite. Shown as sequence length over course of time. The variation in sequence length over time for diffusion models pose unique system constraints for these UNet-based models. In particular, measured values of memory bandwidth utilization and last level cache (LLC) miss rate over the identical region of interest for Imagen and Parti models reveal they correlate with sequence length variation. Note sequence length of Stable Diffusion model actually goes up to 4096, but not shown here for plotting purposes.}
    \label{fig:seq_len_profiling}
\end{figure*}

Figure \ref{fig:seq_len_profiling} shows sequence length profiled over the course of an inference call for various TTI models. Each point on the x-axis represents a distinct time the Attention module is called during inference, while the y-axis records corresponding sequence length. Each graph is truncated to show each model's fundamental period, or the minimum repeating pattern. Note that for Parti, the sequence length linearly increases over the course of inference. This is due to the autoregessive nature of generation, where each generated token becomes part of the prompt for the next token generation. Muse uses parallel decoding instead of autoregessive generation, which is why the sequence length is constant.

\begin{figure}
     \centering
    
         \includegraphics[width=\linewidth]{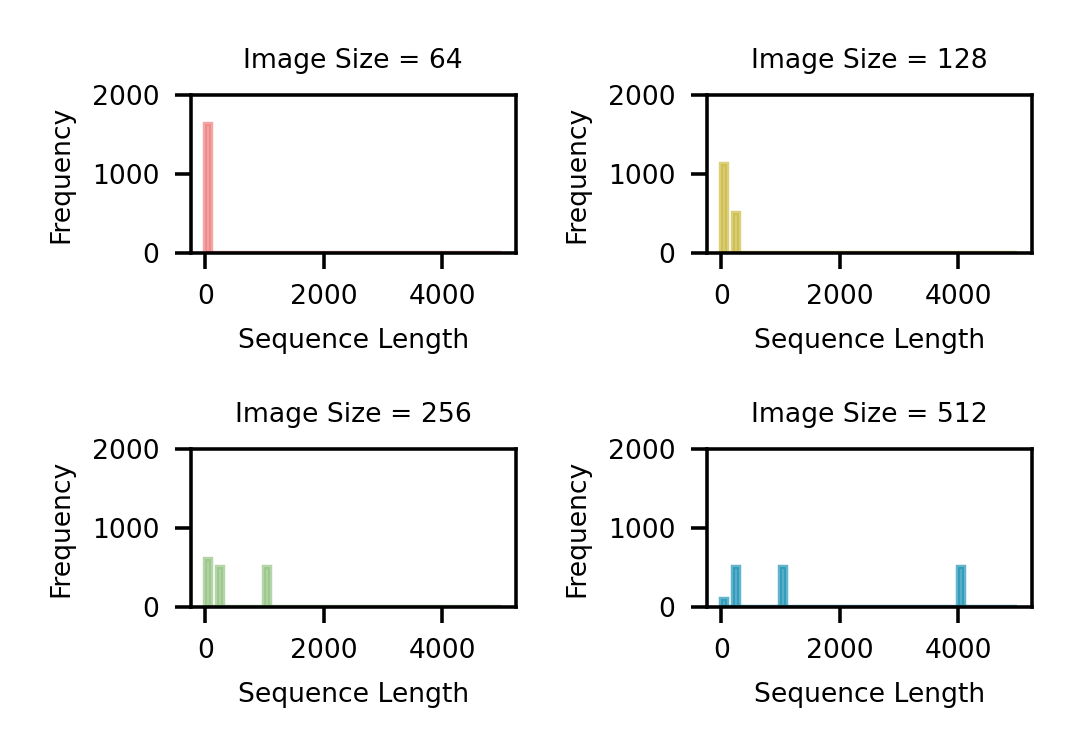}
         
     \caption{Frequency distribution of sequence lengths over the course of inference for Stable Diffusion model. The value distribution shifts right for increasing image size, which corresponds to the input/output size. The distribution associated with image size of 512 corresponds to Figure \ref{fig:seq_len_profiling}.  }
    \label{fig:seq_len_meaning2}
\end{figure}
In contrast, sequence lengths in Diffusion models are (i) \textit{highly variable}, often exhibiting a cyclical pattern, and (ii) can be up to an \textit{order of magnitude smaller} than corresponding LLMs. Note that the U-shaped nature of the sequence lengths are a result of the upsampling/downsampling blocks in the diffusion model, which change the size of the image representation (Figure \ref{fig:detailed_architectures}). The width of the UNet shape is governed by the number of the convolutional blocks in the corresponding model architecture, while the height depends on the starting sequence length governed by desired output image size. 

Note that we additionally quantify the memory bandwidth utilization and last level cache (LLC) miss rate for Imagen and Parti models over the same ROI as Figure \ref{fig:seq_len_profiling}. We find that memory bandwidth utilization and LLC miss rate follow the same pattern as the illustrated sequence length. This highlights the significance of sequence length variation over the course of model inference and its impact on the memory system.

We develop an analytical framework to model the changing memory and FLOPs requirements over the course of the forward pass of a Diffusion model. The model is built by statically examining the model computation to analyze the theoretical memory requirement. We then perform additional profiling runs in order to ensure accuracy of the model, and quantify matrix dimensions throughout model execution to verify the size of the matrices that our formulas predict. 

We start with the desired image size $ H_O \times W_O$, which is subsequently downsampled to latent space,   $ H_L \times W_L $. We define $text\_encode$ as the length of the encoded text prompt. Note that sequence length squared for the \textit{Self-Attention} blocks in the UNet is based on the size of the latent-representation, and is governed by:

\begin{displaymath}
(H_L\cdot W_L)\cdot (H_L\cdot W_L)
\end{displaymath}

while sequence length squared for the \textit{Cross-Attention} blocks is additionally based on text encoding as shown: 

\begin{displaymath}
H_L \cdot W_L \cdot (text\_encode)
\end{displaymath}

We model memory required for the similarity matrix of one Attention calculation in the following formula below. Note that we omit batch size and assume 1 attention head for simplicity. We additionally assume FP16 (i.e. 2 bytes/param).
\begin{displaymath}
2\cdot(H_L \cdot W_L)\cdot (H_L\cdot W_L) + 2\cdot (H_L\cdot W_L)\cdot (text\_encode)
\end{displaymath}

\begin{displaymath}
2 H_L W_L \bigr[ H_L W_L + text\_encode \bigr]
\end{displaymath}

To capture the impact of sequence length variation that comes from the downsampling/upsampling convolutional blocks in the UNet, we model the downsampling factor as $d^n$, where d is the factor by which the image/latent representation is reduced, and $n$ is the diffusion stage. The sequence length of a particular attention calculation depends on the stage of the UNet and the corresponding downsampling factor. This creates the periodic nature observed in Figure \ref{fig:seq_len_profiling}. We then have the following formula for cumulative memory requirements of the similarity matrix over the course of training:

\begin{displaymath}
2  \cdot \Biggr[ 2\sum_{n=0}^{unetdepth-1} \frac{H_L W_L}{d^{2n}} \Bigr[ \frac{H_L W_L}{d^{2n}} + text\_encode \Bigr] 
\end{displaymath}
\begin{displaymath}
+ \frac{H_L W_L}{d^{{2unetdepth}}} \Bigr [ \frac{H_L W_L}{d^{{2unetdepth}}} + text\_encode \Bigr] \Biggr]
\end{displaymath}

Our analysis reveals that since sequence length varies over the course of inference for these Diffusion-based models, memory requirements do as well. In fact, there is a quadratic relationship between latent encoding (or more generally, image size) and sequence length. Then, given the fact that memory required for the similarity matrix scales exponentially with sequence length, the relationship between latent encoding and memory is quadruple. \textbf{As system designers, we must be aware that increasing image resolution or latent size has a \math O(L^4) \endmath   relationship to memory requirements. } This presents challenges, especially for super-resolution networks that operate on high resolutions. These models often  modify the UNet architecture to remove Attention since memory requirements become too large.

The relationship between sequence length and memory requirements leads to potential system optimizations. For example, different denoising steps of the diffusion process could be staggered to allow for maximum memory bandwidth utilization at any one time. Although denoising steps are traditionally sequential, certain steps could potentially be grouped together into pods to allow for this system optimization without significantly impinging on the denoising flow.

\subsection{Implications of Scaling Image Size}

\begin{figure}
     \centering
        \includegraphics[width=\linewidth]{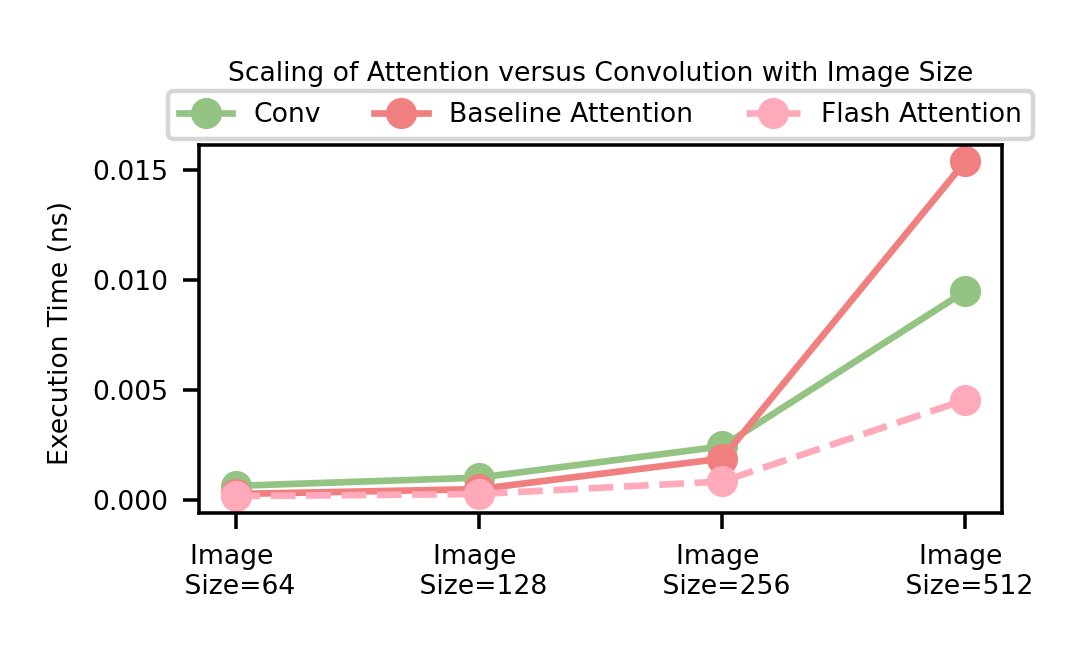}
         
         \caption{Illustration of how time spent on \textit{Attention} versus \textit{Convolution} scales as image size increases for Stable Diffusion. Note that before Flash Attention, \textit{Attention} execution time scales at a faster rate than Convolution execution time with increasing sequence length. However, after Flash Attention is applied, \textit{Convolution} becomes the limiting factor.}
         \label{fig:seq_len_meaning1}
\end{figure}

We subsequently construct a case study on the Stable Diffusion model, in order to more concretely understand the impact of scaling image size (i.e., \math H\_O \endmath, \math W\_O \endmath). Figure \ref{fig:seq_len_meaning2} sweeps different input image sizes, and illustrates the corresponding sequence length distribution for Stable Diffusion inference. Note that as image size increases, sequence length distribution shifts to the right. By examining the case of a 512 x 512 image size, we note the distribution over sequence lengths is relatively equal, which confirms the symmetric nature of the sequence length profile for Stable Diffusion shown in Figure \ref{fig:seq_len_profiling}. As shown, the sequence lengths confine themselves to distinct buckets, which could allow future systems to tailor hardware towards sequence lengths of interest.

Additionally, we augment our \textit{Attention} kernel analysis by comparing to the way in which \textit{Convolution} kernels scale with image size.  We again use the Stable Diffusion model as a representative case study, and sweep the desired output image size from 64 x 64 to 512 x 512, recording \textit{Total Execution Time}. The resulting graph is shown in Figure \ref{fig:seq_len_meaning1}. \textbf{We find that after techniques such as Flash Attention are applied, \textit{Convolution} actually has a larger scaling dependence with image size than \textit{Attention}}.

\section{System Implications of Temporal Dimension} \label{section6}

\begin{figure}
         \centering
        
        \includegraphics[width=0.4\textwidth]{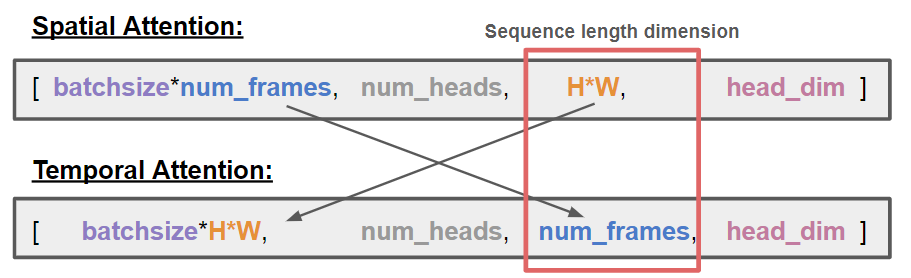}
         \caption{Tensor dimensions are rearranged to perform Spatial versus Temporal Attention. As shown, sequence length is proportional to image size in Spatial Attention and number of frames in Temporal Attention.}

         \label{fig:dim}
\end{figure}

\begin{figure}
     \centering
     \begin{subfigure}[t]{0.48\textwidth}
         \centering

        \includegraphics[width=\textwidth]{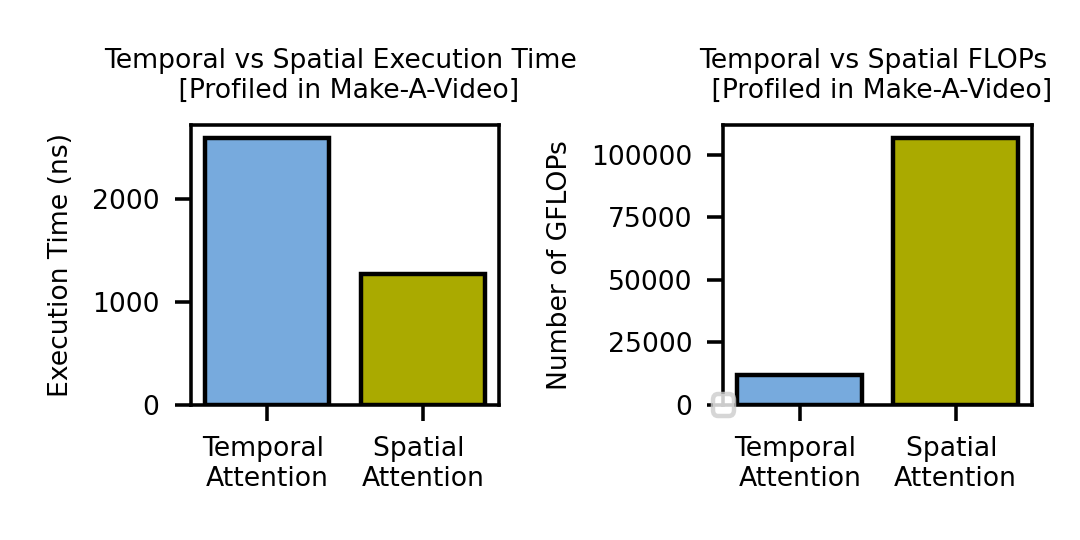}   
     \end{subfigure}%
     \caption{Over the course of Make-A-Video inference, Temporal Attention accounts for 2x the execution time of Spatial Attention, but uses 9x less FLOPs.}
      \label{fig:temporal1}
\end{figure}

\subsection{Temporal versus Spatial Attention}

Text-to-video (TTV) models often augment their TTI backbones with temporal attention in order to connect individually-generated frames from text-to-image models. We find that this \textit{Temporal Attention} exhibits unique system bottlenecks as compared to traditional \textit{Spatial Attention} and explore these implications on future system design. 

As shown in Figure \ref{fig:dim}, the dimensions of the Q/K/V matrices the Attention calculation \cite{flashattention} are reordered for the Temporal Attention, with the desired dimension to attend over shifted into the sequence length position, while other dimensions are shifted into batch size. Specifically, in Temporal Attention, the equivalent sequence length becomes the number of frames of video generation, as the goal of temporal attention is to create a cohesive video in time. In contrast, the sequence length of Spatial Attention is governed by image size.

We subsequently use the Make-A-Video model as a case study for TTV workloads, and profile the Temporal and Spatial Attention modules. We find that \textbf{Temporal Attention takes 2x the execution time of Spatial Attention (Figure \ref{fig:temporal1}a), despite requiring 9x fewer FLOPs (Figure \ref{fig:temporal1}b) when profiled in Make-A-Video.} This suggests that there exists a unique bottleneck within Temporal Attention causing this significantly longer execution time despite fewer computation.

\begin{figure}
     \centering
     \begin{subfigure}[t]{0.48\textwidth}
         \centering

        \includegraphics[width=\textwidth]{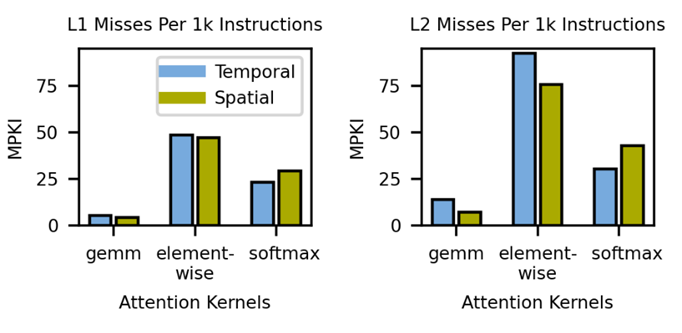}

     \end{subfigure}%
     \hfill
     \begin{subfigure}[t]{0.48\textwidth}
         \centering
         \includegraphics[width=\textwidth]{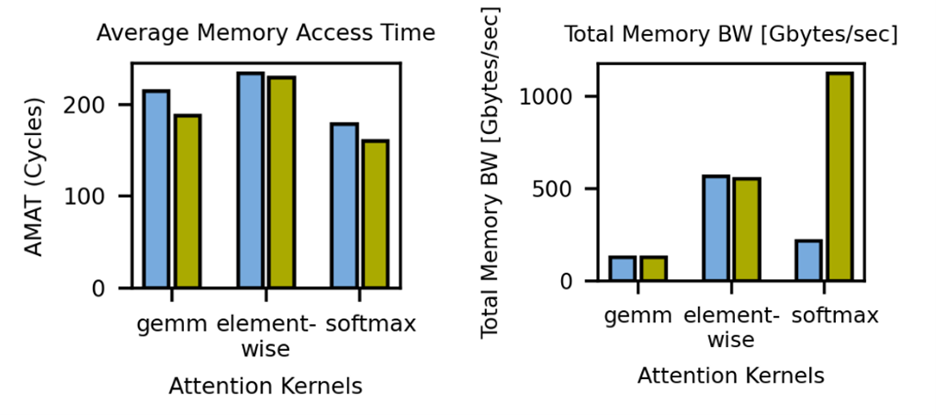}

     \end{subfigure}
     \caption{Performance counters Misses Per Thousand Instructions (MPKI) for L1 and L2 Cache, Average Memory Access Time (AMAT), and Total Memory BW across Temporal and Spatial Attention kernels. Temporal Attention has 1.2-1.96x higher L2 MPKI than Spatial Attention for elementwise and gemm kernels.}

    \label{fig:attn_cache}
\end{figure}

We profile a variety of performance counters to further compare Temporal and Spatial Attention, including Misses Per Thousand Instructions (MPKI), Average Memory Access Time (AMAT), and Memory Bandwidth. As Figure \ref{fig:attn_cache} shows, we find that \textit{GEMM} and \textit{elementwise} kernels have a 1.96x and 1.22x higher L2 MPKI for Temporal attention as compared to Spatial Attention. We additionally find the AMAT for \textit{GEMM}, \textit{elementwise}, and \textit{softmax} kernels is 28, 4, and 18 cycles longer for Temporal Attention, respectively. By collecting an ensemble of performance counters, we make progress towards understanding this system bottleneck, yet there is more work to be done to fully understand the nature of Temporal Attention.

\subsection{Trends in Temporal Attention}

\begin{figure}
     \centering

        \includegraphics[width=\linewidth]{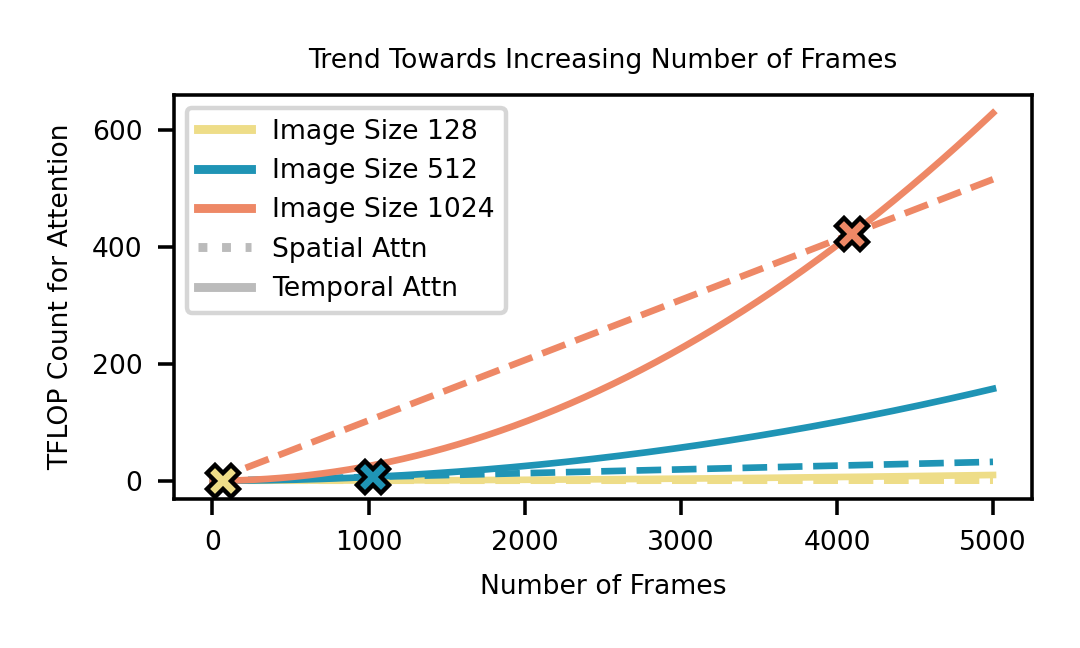}

     \caption{Benchmark illustrating how Temporal Attention FLOPs scale quadratically with number of frames as opposed to Spatial Attention, which scales linearly.}
     \label{fig:temporal_benchmark}
\end{figure}

Given that Temporal Attention is a system performance bottleneck, we create a benchmark based on \cite{timesformer} to project future compute requirements. Figure \ref{fig:temporal_benchmark} sweeps the number of frames (x-axis) and measures the corresponding FLOP count (y-axis), which is calculated by the two main matmul operations in Attention for simplicity. While FLOP count of \textit{Spatial Attention} scales linearly with increasing number of frames, \textit{Temporal Attention} FLOP count scales in polynomial time with equation \(y=AX^2 \), due to frame count being the effective sequence length. For small frame counts, \textit{Temporal Attention} is less computationally expensive than \textit{Spatial Attention}. However, as the number of frames increases, \textit{Temporal Attention} becomes the dominating system bottleneck. Note that increasing image resolution prolongs the cross-over point, as \textit{Spatial Attention} computation is highly impacted by image resolution.

We conclude by outlining that in order to design efficient systems for TTV models, we must anticipate trends towards (i) more frames, and (ii) higher resolutions. First, current videos are typically only seconds in length, with MakeAVideo generating clips of 2-3 seconds only. While frame interpolation can help extend the length of video with a given number of core frames, the generation of movies will require significantly more unique frames to represent distinct concepts. Second, we see a trend towards higher resolutions. Current TTV models stop using Attention when processing high resolutions as it is too memory intensive. This motivates the need for TTV system optimizations that enable long, coherent video generation. 

\section{Related Work}
Recent work has focused on characterizing the system characteristics of LLMs, and in particular, the transformer block.  \cite{efficienttransformerinference} analyzes the latency and model FLOPS utilization (MFU) to understand how to do efficient transformer inference. \cite{bertsystems} analyzes the heterogeneity of GEMM computations in BERT. Flash Attention \cite{flashattention} proposes a tiling technique to reduce HBM accesses and thus improve the \textit{Attention} bottleneck. Flash Attention V2 \cite{flashattention2} and Flash Decoding \cite{flashd} futher introduce more parallelism and inference optimizations, respectively.

Another class of work has focused on optimizing TTI/TTV models from an algorithm perspective. [3] provides an overview of recent progress in diffusion model research, and focuses on evaluating the computational efficiency of these models. [4] proposes a Multi-Architecture Multi-Expert diffusion model in order to better tailor different steps of the diffusion process to their functionality. Others propose using Retrieval Augment Generation (RAG) techniques to supplement the model architectures \cite{racm3}. 

\section{Conclusion}
In this work, we present a detailed system characterization of an emerging class of multi-modal workloads. We find that Diffusion-based TTI models exhibit unique system bottlenecks such as \textit{Convolution} after Flash Attention is applied. We additionally find that unlike LLMs, sequence length changes throughout the course of inference for Diffusion models, complicating the need to design systems for optimal sequence lengths rather than the largest sequence length possible. Finally, we investigate TTV models and find that temporal attention will likely become an increasing bottleneck as we mature in TTV generation. Through our in-depth system performance
analysis, we take a critical first step towards designing efficient
and deployable systems for emerging TTI/TTV workloads.

\section*{Acknowledgements}
This research is not possible without the following colleagues at Meta. We would like to thank Uriel Singer, Adam Polyak, Yipin Zhou to understand the model requirements for text-to-image and large language models and the code base, Driss Guessous for PyTorch’s Scaled Dot Product Attention implementation, and Henry Estela for his support with the training infrastructure.

\bibliographystyle{IEEEtran}
\bibliography{main}

\end{document}